\documentclass[aps, prl, showpacs, twocolumn]{revtex4-1}
\usepackage{amsmath}
\usepackage{amssymb}
\usepackage{graphicx}

\begin{document}

\title{Optimization of the current pulse for spin-torque switches}
\author{Tom Dunn$^1$, and  Alex Kamenev$^{1,2}$ \\
 $^1$Department of Physics, University of Minnesota, Minneapolis, Minnesota 55455, USA. \\
 $^2$Fine Theoretical Physics Institute, University of Minnesota, Minneapolis, Minnesota 55455, USA.}

\begin{abstract}

We address optimization of the spin current intensity profile needed to achieve  spin torque switching of a nanomagnet. For systems with Ohmic
dissipation we prove that the optimal current drives the magnetization along the trajectory, which is exact time-reversed replica of
the relaxation trajectory towards the equilibrium. In practice it means that the optimal current is very nearly {\em twice} the minimal critical
current needed to switch the magnet. Pulse duration of such an optimal current is a slow logarithmic  function of temperature and the
required probability of switching.

\end{abstract}

\pacs{75.70.-i, 85.75.-d, 75.75.Jn}
\maketitle

The spin torque effect, proposed by Slonczewski\cite{Slon96} and Berger\cite{Berger}, is a subject of intense study \cite{Myers99,Katine00,Jonietz2010,Georges2009, Foros2009,grollier:3663,Slavin2009,grollier:509,Chudnovskiy08,Zhang03}, because of its ability to cause magnetic switching in ferromagnetic structures. This opens the possibility of using it for high capacity and low volatility data storage applications. One of the key problems is minimization of the energy dissipation during the switching event. Most of the research centers around optimizing material and geometric parameters of the nanomagnetic structures however some look at how the shape of the spin current pulse can be used \cite{Cui2008,Nikonov2010}. Here we focus
on a much simpler aspect: optimizing intensity and time dependence of the spin current pulse. A very high current density,
while achieving fast switching, results in large Joule heat. On the other hand, the low current density requires a long switching time, which again brings substantial heating. Clearly there is an optimal switching protocol, with minimal deposited heat.

Surprisingly there is a generic prescription for such an optimal current protocol, which optimizes Joule losses irrespectively
to  specific parameters of the structure. We show here that the optimal spin current is such that the sample magnetization retraces
its relaxation trajectory in {\em reversed time}. The relaxation trajectory is the magnetization history starting from the unstable
energy maximum and relaxing to the stable equilibrium in the {\em absence} of any external spin current. In practice it implies that
the optimal current intensity is almost time-independent, which is about {\em twice} the minimal critical switching current.
The duration of such an optimal current pulse is non-universal and is sensitive to the confidence level
of the switching probability as well as temperature.

To be specific we focus on a mono-domain soft ferromagnetic layer, whose magnetization, $\mathbf{M}(t)$, dynamics is described by
the Landau-Lifshitz-Gilbert equation
\begin{eqnarray}
\label{m-det}
\mathbf{\dot{M}} = \mathbf{\dot{M}}_{cons} + \mathbf{\dot{M}}_{diss} + \mathbf{\dot{M}}_{st}\,,
\end{eqnarray}
where $\mathbf{\dot{M}}_{cons}$ corresponds to conservative motion along the Stoner-Wohlfarth (SW) orbitals. The dissipative Gilbert term   $\mathbf{\dot{M}}_{diss}$  is  perpendicular to SW orbitals and causes the magnetization to relax towards the easy axis/external field equilibrium direction.  The last term, $\mathbf{\dot{M}}_{st}$, describes the spin torque effect and has components both perpendicular and parallel to SW orbitals. These three torques may be written as
\begin{eqnarray}
\label{eq:torque}
& &\mathbf{\dot{M}}_{cons} = -\gamma \mathbf{M}\times \mathbf{H}_{eff} \, ,\nonumber \\
& &\mathbf{\dot{M}}_{diss} = -\gamma \alpha \mathbf{\hat{m}}\times\left(\mathbf{M}\times \mathbf{H}_{eff}\right)\,,  \\
& &\mathbf{\dot{M}}_{st} = -\gamma J(t) \mathbf{M}\times\left(\mathbf{\hat{m}}\times \mathbf{\hat{z}}\right)\,. \nonumber
\end{eqnarray}
Here $\mathbf{H}_{eff} = -\nabla_M E/\mu_0$ is the effective magnetic field of the system and $E$ is the energy given by
\begin{eqnarray}
\label{eq:energy}
\frac{E}{\mu_0} = -\frac{H^z_k}{2 M_s} \left( \mathbf{M} \cdot \mathbf{\hat{z}} \right)^2 +\frac{H^x_k}{2 M_s} \left( \mathbf{M} \cdot \mathbf{\hat{x}} \right)^2 - \mathbf{H}_{ext} \cdot \mathbf{M}
\end{eqnarray}
where $\mathbf{M}=\mathbf{\hat{m}}M_s$ and $M_s$ is the saturation magnetization, $H^z_k$ is the easy axis anisotropy field strength with easy axis along the $\mathbf{\hat{z}}$ direction and $H^x_k$ is the strength of the easy $z-y$ plane anisotropy field.
The Gilbert damping constant is $\alpha$ and
$J(t)$ represents the strength of the spin current, which is polarized along the $\mathbf{\hat{z}}$ direction.

\begin{figure}[h]
  \begin{centering}
  \includegraphics[width=8cm]{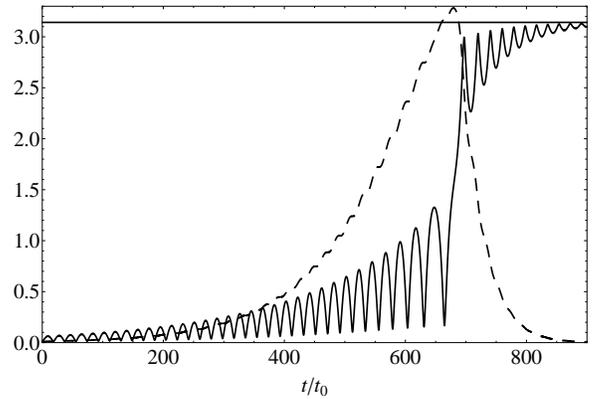}
  \par\end{centering}
  \caption{Azimuthal angle $\theta$ vs. time (full line) and energy vs. time (dashed line) for magnetization reversal of the system with
  $\alpha = 0.01$, $H^z_k = 0.028 M_s$, $H^x_k = M_s$, and $J = 0.01 M_s$. Time is measured in units of $t_0 = \left( \gamma M_s \right)^{-1}$, while the energy in units of $ E_0 / \pi = \mu_0 H_k^z M_s /2\pi$.}\label{thetavstime}
\end{figure}

To analyze the system's dynamics it is convenient to exploit the time scales separation between the precessional time and the switching time.  Figure \ref{thetavstime} shows the azimuthal angle evolution during the switching process. The system goes through many revolutions as it moves out of its initial $\theta=0$ direction towards the switching point at $\theta=\pi/2$, and further during its relaxation towards $\theta=\pi$.  At the same time the energy $E(t)$, Eq.~(\ref{eq:energy}), is a rather smooth function of time, which first increases (thanks to spin-torque) and then relaxes to the new minimum. This observation suggests to integrate out the fast degree of freedom by averaging over each SW orbital, labeled by its energy $E$, Ref.~\cite{ApalkovPRB}. This procedure results in the effective 1d dynamics, with energy as the only ``coordinate''
\begin{eqnarray}
\label{contour-int}
 \dot{E} &=& \frac{\mu_0}{\gamma M_s P(E)} \oint \left( \mathbf{\dot{M}} \times d\mathbf{M} \right)\cdot \mathbf{\hat{m}} \nonumber\\
&=& - \alpha  I_{diss} (E) +  J(t) \mathbf{\hat{m}} \cdot \mathbf{I}_{st} (E)\,, \label{E-det}
\end{eqnarray}
where $P(E) $ is the period of the SW orbital with energy $E$, while $I_{diss} (E)$ and $\mathbf{I}_{st}(E)$ are the contour integrals
along this orbital of the dissipative and spin torques, correspondingly

\begin{eqnarray}
I_{diss} (E) &=&  \frac{\mu_0}{P(E)} \oint \left[d\mathbf{M}\times \mathbf{H}_{eff}\right] \cdot \mathbf{\hat{m}}\,, \\
\mathbf{I}_{st}(E) &=&  \frac{\mu_0}{ M_s P(E)} \oint d\mathbf{M}\times \mathbf{M}\,.
\end{eqnarray}

\begin{figure}[h]
  \begin{centering}
  \includegraphics[width=8cm]{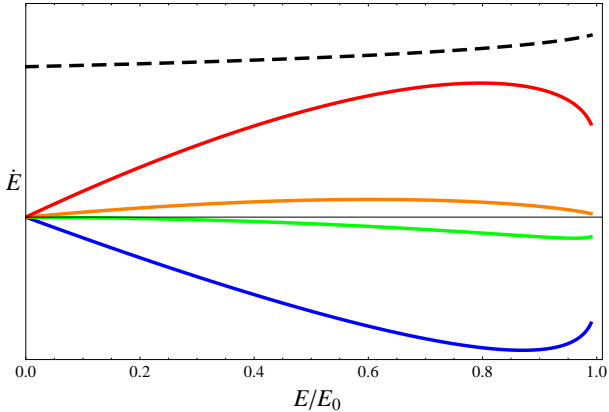}
  \par\end{centering}
  \caption{(Color online) Energy flow $\dot{E}$, Eq.~(\ref{contour-int}), in arbitrary units as a function of energy $E$ for various strengths of the spin current $J$ for the system with $\alpha = 0.01$, $H^z_k = 0.028M_s$ and $H^x_k = M_s$.  From top to bottom $J=0.012M_s$ (red), $J_c=0.0066M_s$ (orange), $J=0.0052M_s$ (green), $J=0$ (blue).  Black dashed line indicates the optimal spin current $J_{opt}$, Eq.~(\ref{efficient-path}), as a function of energy in arbitrary units.}\label{energycurrent}
\end{figure}

Figure~\ref{energycurrent} shows how $\dot{E}$ behaves under various strengths of the applied spin current $J$.
Without the spin current the energy flow is of purely dissipative nature $-\alpha I_{diss} (E)<0$, which forces the magnetization to relax towards $\theta=0$. On the other hand, for the spin current above a certain critical value $J>J_c$, the energy flow is positive, resulting in an eventual magnetization switch, as shown in Fig.~\ref{thetavstime}. There is also a narrow interval of the spin currents $0.82 J_c \lesssim J\lesssim J_c$, where the energy flow is initially positive and changes sign at some energy below the maximum. This leads to stable precession around a proper SW orbital observed in a number of studies \cite{Krivorotov,Rippard04,Kiselev03,Tsoi00}.

In order to find the optimal switching protocol $J(t)$ we require that the Joule heat $\int dt J^2(t) R$, deposited during the switching event, is minimal. Here we assume for simplicity that the total current is proportional to the spin-current $J$. We also assume that the resistance $R$ of the device stays approximately constant as the magnetization direction evolves. 
To ensure the minimum of the Joule heat one needs to require that the ratio of the energy gain $\dot E(t)$ to the energy loss $J^2(t)R$  is maximized at each instance of time. Substituting the energy equation of motion (\ref{contour-int}) and taking derivative with respect to the current, one finds the following condition
\begin{eqnarray}
0 &=& \frac{\partial}{\partial J} \left(-\frac{\alpha I_{diss} (E)}{J^2R} + \frac{\mathbf{\hat{m}} \cdot \mathbf{I}_{st} (E)}{JR} \right) \nonumber\\
&=& 2\frac{\alpha I_{diss} (E)}{J^3R} - \frac{\mathbf{\hat{m}} \cdot \mathbf{I}_{st} (E)}{J^2R} \,.
\end{eqnarray}
Solving for the spin current $J$, yields
\begin{eqnarray}
J_{opt} = 2\alpha\, \frac{I_{diss} (E)}{\mathbf{\hat{m}} \cdot \mathbf{I}_{st}(E)}\, .
\label{efficient-path}
\end{eqnarray}
Therefore the optimal current is exactly {\em twice} the one needed to nullify the energy flow  (\ref{contour-int}).
Substituting this back into equation (\ref{contour-int}), one finds that the latter takes the form
\begin{eqnarray}
\label{eq:time-reversed}
\dot{E} = + \alpha I_{diss} (E)\,.
\end{eqnarray}
This is exactly the same as in Eq.~(\ref{contour-int}) {\em without} the external spin current $J$, but with the time being {\em reversed}.
\begin{figure}[h]
  \begin{centering}
  \includegraphics[width=8cm]{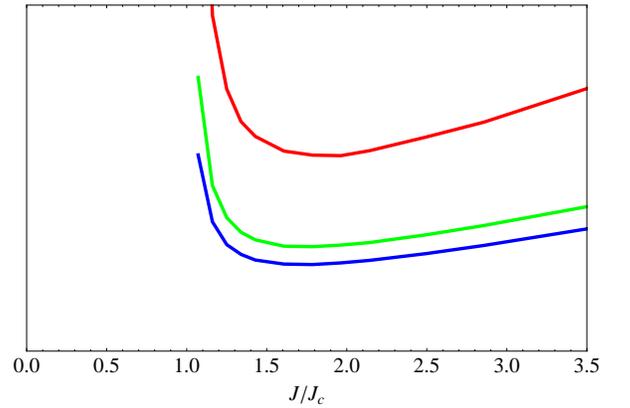}
  \par\end{centering}
  \caption{(Color online) Joule heating during the switching process in arbitrary units vs. spin current for various confidence levels $p$. From top to bottom $p=0.99$ (red), $p=0.5$ (green), $p=0.2$ (blue). Here $\alpha = 0.01$, $H^z_k = 0.028M_s$, $H^x_k = M_s$ and $E_{0}=140 k_b T$.}\label{switching-energy}
\end{figure}
Therefore the optimal current protocol is such that it reverses the purely relaxational trajectory in energy of an isolated system. Such a relaxation trajectory must be thought of as starting at the energy maximum (i.e. the switching point) and winding down towards the minimum at $\theta=0$. Since the ratio of the dissipative and the spin-torque currents along such a trajectory is roughly energy independent, see Fig.~\ref{energycurrent}, equation (\ref{efficient-path}) implies that the optimal current is very nearly a constant given by {\em twice} the critical current $J_c$. This is shown in Fig.~\ref{switching-energy} via Monte-Carlo simulations of the Joule heating as a function of spin current for various confidence levels. Thermal fluctuations enter as a random component $\mathbf{H}_{rand}$ of the external magnetic field with zero average and mean square value $\langle \mathbf{H}_{rand}^2 \rangle = 2\gamma M_s \alpha k_BT \Delta t/\mu_0$ \cite{Brown63} (we assume temperatures large enough to ignore shot noise \cite{Chudnovskiy08}). At low confidence levels $J_{opt} \lesssim 2 J_c$, however at confidence levels nearing $100 \%$ we see the predicted $J_{opt} \approx 2 J_c$.

\begin{figure}[h]
  \begin{centering}
  \includegraphics[width=8cm]{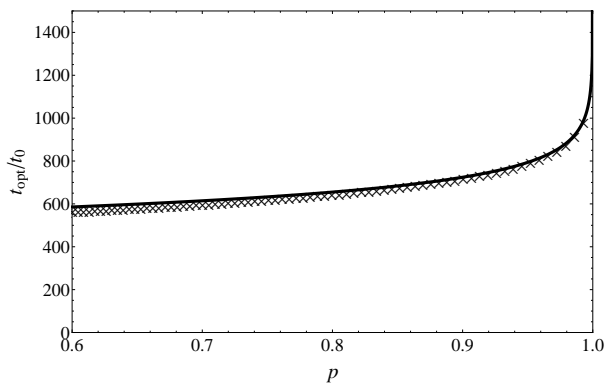}
  \par\end{centering}
  \caption{Simulated optimal switching time as a function of the confidence level (crosses), Eq.~(\ref{eq:optimal-time}) (full line). Same parameters as in Fig.~\ref{switching-energy} with $J = J_{opt}$.}\label{switching-time}
\end{figure}

While we have established that the optimal current pulse is 
$J_{opt}\approx 2J_c$, we have not discussed yet its duration. Here we run into a problem. Indeed, the dissipative relaxation trajectory formally takes an infinite time to approach the energy minimum at $\theta=0$. Thus  its time-reversed, the optimal trajectory, is infinitely long too. What saves the day is the fact that the initial direction of the magnetization is not exactly aligned along $\theta=0$. Instead it is scattered around this direction according to the equilibrium thermal distribution $\propto e^{-E_{ini}/k_BT}$, where $E_{ini}$ is the starting energy counted from its minimal value at $\theta=0$.  Thus the times needed to reach the switching point, which are $\propto \alpha^{-1} \ln (E_0/E_{ini})$ where $E_{0}=\mu_0 H_k^zM_s/2$ is the hight of the energy barrier, are also scattered accordingly.    As a result, any pulse of a finite duration achieves switching only with a certain probability $p<1$. If this probability, i.e. error tolerance, is specified such that $1-p\ll 1$, all realizations with $E_{ini}\geq k_BT(1-p)$ should undergo the switch. This dictates that the duration of the optimal current pulse scales as
\begin{equation}\label{eq:optimal-time}
    t_{opt} \approx \frac{t_0}{ \alpha}\,  \ln \left(\frac{E_{0}}{k_B T(1-p)}\right)\,,
\end{equation}
where $t_0 = \left( \gamma M_s \right)^{-1}$. This result is shown in Fig.~\ref{switching-time} for higher switching probabilities and is in a good agreement with the simulated data.

The statement that the time-reversed trajectory optimizes the switching process is not restricted to the averaging procedure employed in this paper. One can imagine sufficiently complex switching protocols, utilizing the precessional frequency to achieve such an optimal path. For example, Ref.~\cite{Cui2008} used current pulses resonant with the orbital period to excite the system. In the realm of relatively low frequency protocols however, twice the critical current pulse provides the best approximation to the optimal strategy.  The duration of this pulse weakly depends on the confidence level and can be estimated according to Eq.~(\ref{eq:optimal-time}).

We are grateful to M.  Dykman and B. Shklovskii for stimulating discussions. This research was supported  by NSF grant DMR-0804266.

\end{document}